\documentclass[aps,prd,showkeys,showpacs,amssymb,cite,
amsfonts,epsf,preprintnumbers,nofootinbib,superscriptaddress]{revtex4}

\usepackage[dvips]{graphicx}
\usepackage{bm,latexsym,amsmath,amssymb,amsfonts,color}

\newcommand{\be}{\begin{equation}}
\newcommand{\ee}{\end{equation}}
\newcommand{\bear}{\begin{eqnarray}}
\newcommand{\eear}{\end{eqnarray}}
\newcommand{\ba}{\begin{array}}
\newcommand{\ea}{\end{array}}
\newcommand{\nn}{\nonumber}

\begin{document}

\title{Primordial Power Spectra of EiBI Inflation in Strong Gravity Limit}

\author{Inyong Cho}
\email{iycho@seoultech.ac.kr}
\affiliation{Astroparticule et Cosmologie, Universit\'e Paris Diderot, 75013 Paris, France}
\affiliation{Institute of Convergence Fundamental Studies \& School of Liberal Arts,
Seoul National University of Science and Technology, Seoul 139-743, Korea}
\author{Naveen K. Singh}
\email{naveen.nkumars@gmail.com}
\affiliation{Centre for Theoretical Physics, Jamia Millia Islamia, New Delhi-110025, India}

\begin{abstract}
We investigate the scalar and the tensor perturbations of the $\varphi^2$ inflation model
in the strong-gravity limit of Eddington-inspired Born-Infeld (EiBI) theory.
In order to consider the strong EiBI-gravity effect,
we take the value of $\kappa$ large, where $\kappa$ is the EiBI theory parameter.
The energy density of the Universe at the early stage is very high,
and the Universe is in a strong-gravity regime.
Therefore, the perturbation feature is not altered from what was investigated earlier.
At the attractor inflationary stage, however, the feature is changed
in the strong EiBI-gravity limit.
The correction to the scalar perturbation in this limit comes mainly via
the background matter field,
while that to the tensor perturbation comes directly from the gravity ($\kappa$) effect.
The change in the value of the scalar spectrum is little
compared with that in the weak EiBI-gravity limit, or in GR.
The form of the tensor spectrum is the same with that in the weak limit,
but the value of the spectrum can be suppressed down to zero in the strong limit.
Therefore, the resulting tensor-to-scalar ratio can also be suppressed in the same way,
which makes $\varphi^2$ model in EiBI theory viable.
\end{abstract}
\maketitle

\section{Introduction}
In modern cosmology, inflation is regarded as a successful scenario
in explaining the early stage of the Universe.
There are about 190 inflation models suggested for the last 35 years.
Differences originate from varying the matter sector and the gravity sector.

Inflationary models can now be tested by precise modern astronomical observations.
A number of series satellite observations
for Cosmic Microwave Background Radiation (CMBR) have been performed such as WMAP, PLANCK, etc.
The most recent observation by PLANCK ruled out many of inflation models.
The key judgements would be the tensor-to-scalar ratio ($r$) and the spectral index ($n_s$).
The tensor-to-scalar ratio, in particular, has attracted much attention for the past one year,
owing to the ground-base BICEP2 observation \cite{Ade:2014xna}.
The result asserted that the large tensor-mode contribution was observed.
However, the very recent analyses by the PLANCK collaboration \cite{Ade:2015tva,Adam:2015rua}
say that it could be from the dust contamination.
As a result, the value of $r$ needs to be suppressed in many of the models.

The chaotic inflation driven by a massive scalar field has been regarded as
the most standard form of the inflation model \cite{Linde:1983gd}.
However, the value of the tensor-to-scalar ratio predicted by this mode is $r\sim 0.131$
which is beyond the bound of the recent observation.
Although this model provides a very successful fit for the value of $n_s$,
it requires an improvement in the tensor-to-scalar ratio.
Very recently, this model was developed in a different approach in the gravity sector
based on the so-called Eddington-inspired Born-Infeld (EiBI) gravity~\cite{Cho:2013pea}.

The EiBI theory of gravity is described by the action~\cite{Banados:2010ix},
\begin{eqnarray}\label{action}
S_{{\rm EiBI}}=\frac{1}{\kappa}\int
d^4x\Big[~\sqrt{-|g_{\mu\nu}+\kappa
R_{\mu\nu}(\Gamma)|}-\lambda\sqrt{-|g_{\mu\nu}|}~\Big]+S_{\rm M}(g,\varphi),
\end{eqnarray}
where $\kappa$ is the EiBI theory parameter,
$\lambda$  is a dimensionless parameter which is related with
the cosmological constant by $\Lambda = (\lambda -1)/\kappa$
(in this work, we set $8\pi G=1$).
The theory is based on the Palatini formalism;
the metric $g_{\mu\nu}$ and the connection
$\Gamma_{\mu\nu}^{\rho}$ are treated as independent fields.
The Ricci tensor $R_{\mu\nu}(\Gamma)$ is evaluated solely by the connection.
The matter sector described by $S_{\rm M}(g,\varphi)$ is as usual
which couples only to the gravitational field $g_{\mu\nu}$.

The inflationary model in this theory developed in Ref.~\cite{Cho:2013pea}
uses the same type of the matter action
as for the chaotic inflation model \cite{Linde:1983gd} in general relativity (GR),
\be \label{S:chaotic}
S_{\rm M}(g,\varphi) = \int d^4 x \sqrt{-|g_{\mu\nu}|}
\left[ -\frac12 g_{\mu\nu} \partial^\mu\varphi \partial^\nu \varphi -V(\varphi) \right].
\qquad
V(\varphi) = \frac{m^2}{2} \varphi^2.
\ee

By performing variation of the action \eqref{action}
with respect to the metric $g_{\mu\nu}$ and the connection $\Gamma_{\mu\nu}^{\rho}$,
one can cast two equations of motion in the form of
\begin{align}
\frac{\sqrt{-|q|}}{\sqrt{-|g|}}~q^{\mu\nu}
& =\lambda g^{\mu\nu} -\kappa T^{\mu\nu},\label{eom1}\\
q_{\mu\nu} & = g_{\mu\nu}+\kappa R_{\mu\nu}, \label{eom2}
\end{align}
where $T^{\mu\nu} =(2/\sqrt{-|g|}) \delta L_{\rm M}/\delta g_{\mu\nu}$
is the energy-momentum tensor in the standard form,
and $q_{\mu\nu}$ in Eq.~\eqref{eom2} behaves as an auxiliary metric.
Let us take the metric ansatz as
\begin{align}
g_{\mu\nu}dx^\mu dx^\nu
= -dt^2 +a^2(t) \delta_{ij}dx^idx^j
= a^2(\eta) \left( -d\eta^2 +\delta_{ij}dx^idx^j \right),
\end{align}
where $t$ is the cosmological time and $\eta$ is the conformal time.
The scalar-field equation is then given by
\be\label{Seq}
\hat{\hat{\varphi}}_0+3H\hat{\varphi}_0 + \frac{dV}{d\varphi_0}
=\varphi_0''+2{\cal H}\varphi_0' +a^2 \frac{dV}{d\varphi_0} =0,
\ee
where $\hat{} \equiv d/dt$, $' \equiv \d/d\eta$, $H \equiv \hat{a}/a$,
and ${\cal H} \equiv a'/a$.
The subscript $0$ stands for the unperturbed background field.

Let us summarize the background EiBI inflationary feature
investigated in Ref.~\cite{Cho:2013pea}.
There is the so called maximal pressure state (MPS)
beyond which the theory is not well defined.
(Since the EiBI action is of the square-root type,
there is an upper bound in pressure.)
This MPS is the past attractor of the Universe
from which all the evolution paths originated.
The MPS is unstable under the global perturbation.
Therefore, the Universe evolves to the {\it near-MPS stage}
and then enters into the {\it intermediate stage}
which is followed by an inflationary {\it attractor stage}.

The scalar perturbation in this model was investigated
in the weak EiBI-gravity limit ($\kappa \ll m^{-2}$) in Refs.~\cite{Cho:2014jta,Cho:2014xaa}.
The tensor perturbation was investigated in Ref.~\cite{Cho:2014ija}.
The scalar power spectrum $P_{\cal R}$ in the weak-gravity limit
is suppressed little from the spectrum $P_{\cal R}^{\rm GR}$ in GR
by a factor of $(1-\kappa m^2)^2/(1-4\kappa m^2/3)^{1/2}$.
It also exhibits a peculiar rise for very long-wavelength modes.
The tensor spectrum $P_{\rm T}$ is suppressed from $P_{\rm T}^{\rm GR}$
by a factor of $1/(1+\kappa m^2\varphi_i^2/2)$,
where $\varphi_i$ is the value of the scalar field at the beginning of the attractor stage.
The resulting tensor-to-scalar ratio can then be suppressed
by a considerable amount.
For long-wavelength modes, there also exists the same peculiar rise
as in $P_{\cal R}$.

In this paper, we investigate the scalar and the tensor perturbations
in the strong EiBI-gravity limit ($\kappa \gg m^{-2}$).
We consider that the initial perturbations are produced
at the near-MPS stage and evolve adiabatically
until they reach the attractor stage.
Then the perturbations leave the horizon near the beginning of the attractor stage.
The major difference from the weak-gravity case originates
from the attractor stage
since the near-MPS stage is already in the strong-gravity regime
due to the high-matter density.



This paper is organized as following.
In Sec. 2, we investigate the scalar perturbation in the limit of $\kappa \gg m^{-2}$.
The basic equations are already presented Refs.~~\cite{Cho:2014jta,Cho:2014xaa,Lagos:2013aua},
so we do not reproduce all of them in this paper.
In Sec. 3, we investigate the tensor perturbation in the same limit,
and discuss the tensor-to-scalar ratio.
In Sec. 4, we conclude.

\section{Scalar Perturbation}
In this section, we investigate the scalar perturbation
in the strong-gravity limit ($\kappa m^2 \gg 1$).
The basic equations are the same for the weak-gravity limit ($\kappa m^2 \ll 1$)
investigated in Refs.~\cite{Cho:2014jta,Cho:2014xaa}.
We shall consider the $\Lambda=0$ ($\lambda =1$) case in this paper,
but we keep $\lambda$ in some equations.
We take the ansatz for the auxiliary metric as
\begin{align}
q_{\mu\nu}dx^\mu dx^\nu = b^2(\eta) \left[ -\frac{d\eta^2}{z(\eta)} +\delta_{ij}dx^idx^j \right],
\end{align}
and shall denote $h\equiv b'/b$.
The components of Eq.~\eqref{eom1} give
\be\label{eom1cp}
b^2\sqrt{z}=\left(1+ \kappa\rho_0\right)a^2,
\qquad
\frac{b^2 }{\sqrt{z}}=\left(1 - \kappa p_0\right)a^2,
\ee
where $\rho_0 = \varphi_0'^2/2a^2 + V(\varphi_0)$
and $p_0 = \varphi_0'^2/2a^2 - V(\varphi_0)$.
From Eq.~\eqref{eom1cp}, we get
\be\label{zb}
z=\frac{1+\kappa\rho_0}{1 - \kappa p_0},
\qquad
b = (1+\kappa\rho_0)^{1/4}(1-\kappa p_0)^{1/4}a,
\ee
where $z$ is the key function which controls the scalar perturbation.

When the cosmological constant is absent ($\lambda =1$),
using Eq.\eqref{eom2} one can show that the EiBI action \eqref{action}
is equivalent to a bimetric-like theory action
\be\label{S}
S(g,q,\varphi) = \frac{1}{2} \int d^4x \sqrt{-|q_{\mu\nu}|} \left[  R(q) - \frac{2}{\kappa}  \right]
+{1 \over 2\kappa} \int d^4x \left(\sqrt{-|q_{\mu\nu}|}q^{\alpha\beta}g_{\alpha\beta} - 2\sqrt{-|g_{\mu\nu}|}
\right)+ S_{\rm M}(g,\varphi),
\ee
which we shall use for the scalar perturbation.
The scalar perturbations for the metrics are introduced as
\begin{align}
ds_q^2 &= b^2\left\{-\frac{1+ 2\phi_1}{z} d\eta^2  +2\frac{B_{1,i}}{\sqrt{z}} d\eta dx^i
+ \Big[(1-2\psi_1)\delta_{ij} + 2 E_{1,ij}\Big] dx^i dx^j \right\}, \label{qpert} \\
ds_g^2 &= a^2\left\{-(1+ 2\phi_2) d\eta^2 + 2 B_{2,i}d\eta dx^i
+\Big[(1-2\psi_2)\delta_{ij} + 2 E_{2,ij}\Big]dx^i dx^j \right\},  \label{metric_pert}
\end{align}
and the scalar-field perturbation is given by $\varphi=\varphi_0 +\chi$.
Using these perturbed metrics,
we get the second-order action from Eq.~\eqref{S}.
(For the result, please see Refs.~\cite{Cho:2014jta,Cho:2014xaa}.)

Introducing the Fourier modes for the nine perturbation fields $F_l$ ($l=1\sim 9$) as,
\be\label{Fourier}
F_l(\eta,\vec{x}) = \int \frac{d^3k}{(2\pi)^{3/2}}
F_l(\eta,\vec{k})e^{i\vec{k}\cdot\vec{x}},
\ee
and fixing gauge conditions as
\be\label{gg}
\psi_1=0 \quad{\rm and}\quad E_1=0,
\ee
one can solve all the perturbation fields expressed in terms of $\chi$.
The second-order perturbation action is then written only
by $\chi$ and the background fields as
\be\label{Ss}
S_{\rm s}(\chi) = \frac{1}{2}\int  d^3k d\eta\;
\Big[ f_1(\eta) \chi'^{2} - f_2(\eta,k)  \chi^2 \Big],
\ee
where
\be\label{f12}
f_1(\eta) = \frac{a^2 (3z^2-2z+3)}{(z+1)(3z-1)},
\qquad
f_2(\eta,k) = \frac {\beta}{8\kappa^3 h^2 z^{5/2}(z + 1)^2},
\ee
and $\beta$ is a bit lengthy defined in Refs.~\cite{Cho:2014jta,Cho:2014xaa}.
In addition, we present the perturbation field $\psi_2$ for later use,
\be\label{psi2_XY}
\psi_2 = \frac{z-1}{2\kappa hz(z+1)(3z-1)}
\Big[ -2\kappa hz(z-1){\cal X}\chi' +a^2(z-1)^2{\cal X}\chi +2\kappa hz(3z-1){\cal Y}\chi \Big],
\ee
where
\be\label{XY}
{\cal X} \equiv \frac{1}{a\sqrt{\rho_0+p_0}},\qquad
{\cal Y} \equiv -m\frac{\sqrt{\rho_0-p_0}}{\rho_0+p_0}.
\ee
The matter perturbation $\chi$ is not a canonical field in the action \eqref{Ss}.
We introduce a canonical field $Q$ from the transformations
$Q \equiv \omega\chi$ and $d\tau \equiv (\omega^2/f_1)d\eta$.
The perturbation equation for $Q$ then becomes
\be\label{Q-eq1}
\ddot{Q} + \left(\sigma_s^2 k^2-\frac{\ddot{\omega}}{\omega}\right)Q =0 ,
\ee
where $\dot{} \equiv d/d\tau$ and $\sigma_s^2 \equiv f_1 f_2/k^2\omega^4$.
The new function $\omega$ is determined by assuming
a Bunch-Davies vacuum in the $k\to\infty$ limit.
Requiring $\sigma_s^2 \to 1$ in this limit, we get
\begin{align}\label{omega}
\omega^4= \frac{a^4 (3 z^2-2 z+3)}{z(z+1)(3z-1)}.
\end{align}
The normalization condition for the canonical field is given by
\be\label{norm}
Q\dot Q^*-Q^*\dot Q =i.
\ee

In this set-up, the scalar perturbation in the weak gravity limit ($\kappa m^2 \ll 1$)
had been investigated in Refs.~\cite{Cho:2014jta,Cho:2014xaa}.
The near-MPS stage is already in a strong gravity limit
because the matter density is very high.
We had not assumed $\kappa m^2 \ll 1$ for the near-MPS stage in Refs.~\cite{Cho:2014xaa},
and thus the near-MPS result still holds in the strong-gravity limit ($\kappa m^2 \gg 1$).
(This is also true for the tensor perturbation in the next section.)

The results for the attractor stage in Refs.~\cite{Cho:2014jta,Cho:2014xaa},
however, can be altered in the strong-gravity limit.
Let us investigate the scalar perturbation at the attractor stage
in this limit.
First of all, we need to check the background fields $a$ and $\varphi_0$
in the strong-gravity limit to see if there is any change.
The Friedmann equation was obtained in Ref.~\cite{Cho:2013pea,Scargill:2012kg},
\begin{align}\label{H}
H &= \frac{\hat a}{a} =
    \frac{1}{\left(\lambda/\kappa + V \right)^2 + \hat{\varphi}_0^4/2}
\left\{ -\frac{1}{2}\left(\frac{\lambda}{\kappa}+V + \frac{\hat {\varphi}_0^2}{2}\right)
     V'({\varphi}_0)\hat {\varphi}_0  \right. \nn\\
     &\pm\frac{1}{\sqrt{3}}\left(\frac{\lambda}{\kappa}+V-\frac{\hat {\varphi}_0^2}2\right) \times
     \left. \left[\left(\frac{\lambda}{\kappa}+V +\frac{\hat {\varphi}_0^2}{2}\right)^{3/2}
\left(\frac{\lambda}{\kappa}+V-\frac{\hat {\varphi}_0^2}2\right)^{3/2}
-\frac{1}{\kappa}\left(\frac{\lambda}{\kappa}+V + \frac{\hat {\varphi}_0^2}{2}\right)
    \left( \frac{\lambda}{\kappa}+V -\hat {\varphi}_0^2 \right)
     \right]^{1/2}  \right\}.
\end{align}
With the first slow-roll condition, $\hat{\varphi}_0^2/2 \ll V$,
the term with $V'(\varphi_0)$ (let us denote this term by $H_1$)
and the other term ($H_2$) are approximated by
\be
H_1 \approx -\frac{V'\hat{\varphi}_0}{2(\lambda/\kappa + V)},
\qquad
H_2 \approx \pm \sqrt{\frac{1}{3} \left(V+\frac{\lambda-1}{\kappa}\right)}.
\ee
In both limits of $\kappa m^2 \ll 1$ and $\kappa m^2 \gg 1$,
one can easily check $H_1^2 \ll H_2^2$.
Then the equation is simplified to
\be
H \approx H_2 \approx \pm \sqrt{\frac{1}{3} \left(V+\Lambda\right)},
\ee
which is the Friedmann equation in GR.
(Depending on the value of $\lambda$,
$H_1$ can be larger than the cosmological constant term in $H_2$.
However, when we consider $\Lambda=0$,
the above result is unchanged.)
Then the scalar-field equation \eqref{Seq} is not changed,
and the background solutions for $a$ and $\varphi_0$ are as usual.
Applying the first and the second slow-roll conditions
to the scalar-field and the Friedmann equations, one gets
\be\label{slowsol}
\varphi_0(t) \approx \varphi_i +\sqrt{\frac{2}{3}}m t,
\qquad
a(t) \approx a_i\; e^{[\varphi_i^2-\varphi_0^2(t)]/4},
\ee
where the subscript $i$ stands for the beginning of the attractor stage.
At the early stage of the attractor, $m^2t^2 \ll mt$,
the scale factor is further approximated to that of de Sitter as
\be\label{aATT}
a(t) \approx a_i e^{-\varphi_i mt/\sqrt{6} - m^2t^2/6}
\approx a_i e^{-\varphi_i mt/\sqrt{6}}.
\ee
Using the first slow-roll condition
$\hat{\varphi}_0^2/2 \ll m^2\varphi_0^2/2$ and $\kappa m^2 \gg 1$,
we get from Eq.~\eqref{zb}
\be\label{z_approx}
z= \frac{1+\kappa\rho_0}{1 - \kappa p_0}
= \frac{1+\kappa (\hat{\varphi}_0^2/2 + m^2\varphi_0^2/2)}{1-\kappa (\hat{\varphi}_0^2/2 - m^2\varphi_0^2/2)}
= 1 + \frac{\kappa\hat{\varphi}_0^2}{1-\kappa (\hat{\varphi}_0^2/2 - m^2\varphi_0^2/2)}
\approx 1 + \frac{2\kappa m^2/3}{1+\kappa m^2\varphi_0^2/2}
\approx 1+\frac{4}{3\varphi_0^2},
\ee
and
\be\label{b_approx}
b = (1+\kappa\rho_0)^{1/4}(1-\kappa p_0)^{1/4}a
\approx (1+\kappa\rho_0)^{1/2}a
\approx a\sqrt{1+\frac{1}{2}\kappa m^2\varphi_0^2}
\approx -\frac{\sqrt{\kappa}m}{\sqrt{2}}a\varphi_0.
\ee
For this $b$, assuming $\varphi_i\sim {\cal O}(10)$
and with the de Sitter approximation for $a$,
we have
\be
\frac{\hat{b}}{b}
= \frac{\hat{a}}{a} +\frac{\hat{\varphi}_0}{\varphi_0}
\approx -\frac{m\varphi_i}{\sqrt{6}} +\frac{\sqrt{2}m}{\sqrt{3}\varphi_0}
\approx -\frac{m\varphi_i}{\sqrt{6}} = \frac{\hat{a}}{a}
\qquad\Rightarrow\qquad
h\approx {\cal H}.
\ee
Using the above approximations for $z$ and $b$, we have
\begin{align}
{\cal X}^2 &= \frac{1}{a^2(\rho_0+p_0)}=\frac{\kappa\sqrt{z}}{b^2(z-1)}
\approx \frac{3\sqrt{z}}{2m^2a^2},\label{X_approx}\\
{\cal Y} &=-m \frac{\sqrt{\rho_0-p_0}}{\rho_0+p_0}
\approx \frac{3}{2}\varphi_0z^{1/4}.\label{Y_approx}
\end{align}
For the time transformation, we have
\be
d\tau =\frac{\omega^2}{f_1}d\eta = \frac{\omega^2}{af_1}dt
= \sqrt{\frac{(z+1)(3z-1)}{z(3z^2-2z+3)}}\frac{dt}{a}
\approx \left( 1-\frac{8}{9\varphi_i^4} \right)\frac{dt}{a}
\equiv \frac{1}{\tau_c} \frac{dt}{a}.
\ee
Here, the factor $8/9\varphi_i^4$ is the next (second) order of the correction
that we consider in this work.
Therefore, we shall ignore it and set $\tau_c=1$
in following calculations.
For the scale factor given in Eq.~\eqref{aATT}, we have
\be
a(\tau) \approx \frac{\sqrt{6}}{\varphi_i m(\tau-\tau_0)},
\quad\mbox{and}\quad
\frac{\ddot a}{a} \approx \frac{\varphi_i^2m^2a^2}{3},
\ee
where $\tau_0$ is the moment of the end of inflation.
(We set $\tau =0$ for $t\to-\infty$
which corresponds to the beginning of the Universe at the near-MPS stage.)
From Eq.~\eqref{omega}, we have an approximation,
\be
\omega \approx \left( 1-\frac{2}{3\varphi_0^2} \right) a,
\ee
which gives
\be
\frac{\ddot\omega}{\omega} \approx \frac{\ddot a}{a}
+ \frac{4}{3\varphi_0^2-2} \left[ \frac{\ddot{\varphi}_0}{\varphi_0}
-3\left( \frac{\dot\varphi_0}{\varphi_0} \right)^2
+2 \frac{\dot a\dot\varphi_0}{a\varphi_0}\right]
\approx \frac{\ddot a}{a}
\approx \frac{\varphi_i^2m^2a^2}{3},
\ee
where the square-bracket term is the second or higher order,
so we ignore them again.

Getting the correction for $f_2$ is lengthy.
Let us define small variables as
$\epsilon_1\equiv 1/\varphi_i \sim {\cal O}(10^{-1})$
and $\epsilon_2\equiv 1/\sqrt{\kappa}m < {\cal O}(10^{-1})$,
and keep their second-order terms.
Then we finally get
\be
f_2 \approx \left( 1-\frac{2}{9}\epsilon_1^2 \right)  k^2a^2
- \left[ 1-\epsilon_1^2\left( \frac{14}{9}-8\epsilon_2^2 \right)  \right] m^2a^4,
\ee
and
\be
\sigma_s^2k^2 = \frac{f_1f_2}{\omega^4} =\frac{zf_2}{a^2}
\approx \left( 1+\frac{10}{9}\epsilon_1^2 \right)  k^2
- \left[ 1-\epsilon_1^2\left( \frac{2}{9}-8\epsilon_2^2 \right)  \right] m^2a^2.
\ee
Finally we get
\be
\sigma_s^2k^2 -\frac{\ddot\omega}{\omega}
\approx \left( 1+\frac{10}{9\varphi_i^2} \right)  k^2
- \left[ \frac{\varphi_i^2}{3} +1-\left( \frac{2}{9}-\frac{8}{\kappa m^2} \right) \frac{1}{\varphi_i^2} \right] m^2a^2
\approx k^2 -\frac{\varphi_i^2m^2a^2}{3}
\approx k^2 - \frac{2}{(\tau-\tau_0)^2}.
\ee
Therefore, the strong-gravity effect is minor for the mode equation,
\be\label{Q-eq2}
\ddot{Q} + \left(\sigma_s^2 k^2-\frac{\ddot{\omega}}{\omega}\right)Q
\approx \ddot{Q} + \left[ k^2 - \frac{2}{(\tau-\tau_0)^2} \right]Q
\approx 0,
\ee
and the mode function has the same form with that in the weak-gravity limit and in GR,
\begin{align}
Q_{\rm ATT}(\tau)  &\approx A_1 \left[ \cos k(\tau-\tau_0) - \frac{\sin k(\tau-\tau_0)}{k(\tau-\tau_0)} \right]
+A_2 \left[ \sin k(\tau-\tau_0) + \frac{\cos k(\tau-\tau_0)}{k(\tau-\tau_0)} \right] \label{muATTmt1} \\
&= A_1' \left[1+ \frac{i}{k (\tau-\tau_0)}\right] e^{ik(\tau-\tau_0)}
+A_2'  \left[1- \frac{i}{k (\tau-\tau_0)}\right] e^{-ik(\tau-\tau_0)},\label{muATTmt2}
\end{align}
where $A_1'\equiv (A_1-i A_2)/2$ and $A_2' \equiv (A_1+i A_2)/2$.
The coefficients $A_i'$s are to be determined by imposing the initial conditions
when the perturbation is produced at the near-MPS stage.

At the near-MPS stage for the strong-gravity limit,
as it was mentioned earlier,
we can use the result in Ref.~\cite{Cho:2014xaa}.
It is summarized as follows.
At the near-MPS stage, $z\gg1$ from Eq.~\eqref{zb}.
The time coordinates are transformed as
\be
d\tau =\frac{\omega^2}{f_1}d\eta = \frac{\omega^2}{af_1}dt
\approx \frac{dt}{a\sqrt{z}}
\approx \frac{\sqrt{-\psi_0}}{a_0} e^{\sqrt{2/3\kappa}t} dt
\quad\Rightarrow\quad
\tau \approx \sqrt{-\frac{3\kappa\psi_0}{2a_0^2}} e^{\sqrt{2/3\kappa}t},
\ee
where $\psi_0$ is a small negative free parameter,
and we set $\tau =0$ for $t\to-\infty$.
The perturbation equation is then given by
\be\label{Q-eq3}
\ddot{Q} + \left( k^2 + \frac{1}{4\tau^2} \right)Q \approx 0,
\ee
and the solution is given by
\be\label{Qsol3}
Q_{\rm MPS}(\tau)  \approx \sqrt{\tau}\left[c J_0(k\tau)
+ \left( R-i\frac{\pi}{4c}\right)Y_0(k\tau) \right],
\ee
where the coefficients are to be determined by imposing initial conditions.
In EiBI inflation, there is no obstacle to consider $\tau \to 0$ $(t\to -\infty)$
because the background Universe is regular at past infinity.
For the perturbation, however,
$Y_0(k\tau)$ in the solution \eqref{Qsol3} diverges at $\tau\to 0$.
One may not kill this term by setting the coefficients
because $Q_{\rm MPS}$ does not satisfy the normalization condition \eqref{norm}
solely with the $J_0(k\tau)$ term.

Then how do we manage this problem?
The solution is given by the instability of the MPS state
and the quantum fluctuation of the background scalar field.
As it was investigated in Ref.~\cite{Cho:2013pea},
the Universe at the MPS state is unstable to a global perturbation,
evolves to the near-MPS stage, and then to the intermediate stage.
(Please see the diagrams in Fig. 1 in Ref.~\cite{Cho:2013pea}.)
After the Universe enters the attractor stage,
if it spends an appropriate
amount of time to acquire necessary $e$-foldings,
it will be an ideal situation.
However, it requires very fine tuning at the near-MPS stage.
Most of the evolution paths may leave the near-MPS stage early,
then it enters into the region where $H$ is large in the phase diagram
during the intermediate stage.
There, it experiences large quantum fluctuations as
$\delta\varphi \sim {\cal O}(H)$ and
$\delta\hat\varphi \sim \delta\varphi/\delta t \sim {\cal O}(H^2)$.
Then the state of the Universe is pushed back to the near-MPS region in the diagram.
The Universe repeats this process until its path avoids the large fluctuation (large $H$) region,
and settles down to the attractor stage.
Then it is reasonable to impose the initial condition for the perturbation
in the near-MPS region for the final path.
In addition, to treat the perturbation in a classical way,
the wavelength scale $\lambda_{\rm phys}$ at the production moment ($\tau_*$)
needs to be comparable to the Planck scale $l_{p}$,
\be\label{lplanck}
\lambda_{\rm phys} = \frac{a(\tau_*)}{k} \gtrsim l_p
\qquad\Rightarrow\qquad
\tau_* \gtrsim a^{-1} (kl_p)
\approx \sqrt{-\frac{3\kappa\psi_0}{2a_0^2}}
\left[ \frac{kl_p}{a_0(2\kappa)^{1/3}} \right]^{\sqrt{3/2\kappa m^2}}.
\ee

We impose the minimum-energy condition at $\tau_*$ for the initial perturbation,
then we get the coefficients of the solution \eqref{Qsol3} as
\be\label{cR}
c^2 = \frac{\pi}4 \frac{Y^2 +Y_0^2}{|JY_0-J_0Y|},
\qquad
R=\mp \sqrt{\frac{\pi}{4}}\frac{JY + J_0 Y_0}{\sqrt{|JY_0-J_0Y|(Y^2 +Y_0^2)}},
\ee
where $J \equiv {(J_0-2k\tau_* J_1)}/{\sqrt{1+ 4k^2\tau_*^2}}$,
$Y \equiv {(Y_0-2k\tau_* Y_1)}/{\sqrt{1+ 4k^2\tau_*^2}}$,
$J_{0,1} \equiv J_{0,1}(k\tau_*)$, and $Y_{0,1} \equiv Y_{0,1}(k\tau_*)$.
For high $k$-modes, $R\to 0$ and $c^2 \to\pi/4$
which correspond to the positive-energy mode.
For low $k$-modes, the initial perturbation is a mixture of the
positive- and the negative-energy modes.

The perturbations produced at the near-MPS stage
evolve to the {\it intermediate stage}
which is connected to the inflationary {\it attractor stage}.
The power spectrum is evaluated by the mode function $Q_{\rm ATT}$,
but its coefficients are determined from the initial perturbation $Q_{\rm MPS}$.
In order to express the coefficients $A_i'$s of $Q_{\rm ATT}$
in terms of $c$ and $R$ of $Q_{\rm MPS}$,
we need a solution matching.
However, the mode function at the intermediate stage is not known.
We assume that the perturbation evolves adiabatically,
then the mode solution to the perturbation equation
$\ddot Q + \Omega_k^2(\tau) Q =0$  at the intermediate stage
can be expressed by the WBK approximation,
\be\label{sol:WKB}
Q_{\rm WKB}(\tau) = \frac{b_1}{\sqrt{2\Omega_k(\tau)}}
    \exp\left[ i \int^\tau \Omega_k(\tau') d\tau'\right]
+ \frac{b_2}{\sqrt{2\Omega_k(\tau)}}
    \exp\left[ -i \int^\tau \Omega_k(\tau') d\tau'\right],
\ee
for which the adiabatic condition is
$\Omega_k^{-3}\left| d\Omega_k^2/d\tau \right| \ll 1$.

As it was presented in Ref. \cite{Cho:2014ija},
matching $Q$'s and $\dot Q$'s at $\tau_1$ for MPS and WKB
and at $\tau_2$ for WKB and ATT,
we have
\begin{align}
b_{1,2} &\approx \frac{c_1 \mp i c_2}{\sqrt{\pi}} \,e^{ \pm i (k\tau_1 -\pi/4)},\label{b12}\\
A_{1,2}' &\approx \frac{e^{\mp ik(\tau_2-\tau_0)}}{2}
\left[ Q_{\rm WKB}(\tau_2;b_1,b_2) \mp \frac{i}{k} \dot Q_{\rm WKB}(\tau_2;b_1,b_2) \right]. \label{A12}
\end{align}
At the end of inflation, we have
$Q_{\rm ATT}(\tau) \approx i(A_1'-A_2')/k(\tau-\tau_0)$,
and from Eqs. \eqref{b12} and \eqref{A12},
we get
\be
|Q_{\rm ATT}|^2 \approx
\frac{|A_1'-A_2'|^2}{k^2(\tau-\tau_0)^2}
=  \frac{c^2+R^2 + \pi^2/16c^2}{\pi k^3 (\tau-\tau_0)^2}.
\ee

We consider the perturbations produced at the near-MPS stage
and exiting the horizon near the beginning of the attractor stage
($\varphi_0\sim\varphi_i$).
The power spectrum is evaluated for these modes when they cross the horizon.
In evaluating the power spectrum,
the perturbation field $\psi_2$ contributes to the EiBI correction as below.
From Refs.~\cite{Cho:2014jta,Cho:2014xaa}, we have
\be\label{psi2_XY}
\psi_2 = \frac{z-1}{2\kappa hz(z+1)(3z-1)}
\Big[ -2\kappa hz(z-1){\cal X}\chi' +a^2(z-1)^2{\cal X}\chi +2\kappa hz(3z-1){\cal Y}\chi \Big].
\ee
With the approximations obtained earlier,
each term in the square bracket is approximated as
\begin{align}
-2\kappa hz(z-1){\cal X}\chi' &\approx \frac{4\varphi_i}{3m} \frac{\epsilon_1^2}{\epsilon_2^2} a\chi,\\
a^2(z-1)^2{\cal X}\chi &\approx \frac{8\sqrt{2}}{3\sqrt{3}m} \epsilon_1^4a\chi,\\
2\kappa hz(3z-1){\cal Y}\chi &\approx -\frac{\sqrt{6}\varphi_i}{m}\frac{1}{\epsilon_1\epsilon_2^2} a\chi,
\label{psi2-third}
\end{align}
where we used $\chi'=d\chi/d\eta = ad\chi/dt \sim ma\chi$.
Considering $\varphi_i\sim 1/\epsilon_1$, and $\epsilon_1,\epsilon_2 \ll 1$,
the third term \eqref{psi2-third} is most dominant and we get
\be
\psi_2 = \frac{z-1}{z+1}{\cal Y}\chi
\approx \epsilon_1 \left( 1+\frac{2}{3}\epsilon_1^2 \right)^{-1}
\left( 1+\frac{4}{3}\epsilon_1^2 \right)^{1/4} \chi
\approx \epsilon_1 \left( 1-\frac{1}{3}\epsilon_1^2 \right) \chi
\approx \frac{\chi}{\varphi_0}.
\ee
The comoving curvature becomes then
\be
{\cal R} = \psi_2 +\frac{H}{\hat{\varphi}_0}\chi
\approx \left( \frac{1}{\varphi_0}-\frac{\varphi_i}{2} \right)\chi.
\ee
The scalar power spectrum evaluated at the horizon crossing becomes
\begin{align}
P_{\cal R} &= \frac{k^3}{2\pi^2}{\cal R}^2
\approx \frac{k^3\varphi_i^2}{8\pi^2} \left( 1-\frac{2}{\varphi_i^2} \right)^2
\left| \frac{Q_{\rm ATT}}{\omega_{\rm ATT}} \right|^2 \\
&\approx \frac{2}{\pi}\left( c^2+R^2 + \frac{\pi^2}{16c^2} \right)
\times \left( 1-\frac{2}{\varphi_i^2} \right)^2 \left( 1-\frac{2}{3\varphi_i^2} \right)^{-2}
\times \frac{m^2\varphi_i^4}{96\pi^2} \\
&\equiv D_k \times E_{\varphi_i}^{\rm S} \times P_{\cal R}^{\rm GR}. \label{EkS}
\end{align}
Here, $P_{\cal R}^{\rm GR} \equiv m^2\varphi_i^4/96\pi^2$ is the power spectrum in GR,
and $E_{\varphi_i}^{\rm S} \equiv (1-2/\varphi_i^2)^2/(1-2/3\varphi_i^2)^{-2}$
is the EiBI correction in the strong-gravity limit.
Note that this factor is different from the EiBI correction in the weak-gravity limit,
$E_\kappa^{\rm S} \equiv (1-\kappa m^2)^2/(1-4\kappa m^2/3)^{1/2}$.
In the strong-gravity limit, the first-order correction for the factor $E^{\rm S}$
comes from  the background scalar field $\varphi_0$,
not via $\kappa$ which is the EiBI gravity parameter.
The factor $D_k \equiv (2/\pi) (c^2+R^2 + \pi^2/16c^2)$
is not altered from that in the weak-gravity limit.
$D_k$ is a function of $k$ and $\kappa$ (implied in $\tau_*$).
As it is shown in Fig.~1,
$D_k$ exhibits a peculiar rise at low $k$,
while it becomes unity, $D_k \to 1$, at high $k$.
This peak needs to be within the range of the cosmic variance
in the CMB data, or it should
correspond to the very long-wavelength modes
which are not observable today.

As a summary, compared with the power spectrum in GR,
$P_{\cal R}$ exhibits a peculiar peak for low $k$-modes via $D_k$,
in the same manner as in the weak-gravity limit.
This peculiar feature originates from imposing the initial condition
on $Q_{\rm MPS}$ in Eq.~\eqref{Qsol3}.
[If the perturbation is produced at the attractor stage,
the initial (minimum-energy) condition picks up only the positive-energy mode
of the solution $Q_{\rm ATT}$ in Eq.~\eqref{muATTmt2}.
This makes $D_k = 1$ and there is no peak in the spectrum.]
However, it is evident only for the low $k$-modes (very long-wavelength modes)
which may reside only in the super-horizon scales.
The EiBI correction factor for over-all scales is small,
$E_{\varphi_i}^{\rm S} \sim 1-8/3\varphi_i^2 \lesssim 1$,
and is different from that in the weak-gravity limit.
This correction originates from $\psi_2$ and $\omega$.

\section{Tensor Perturbation and Tensor-to-Scalar Ratio}
In this section, we investigate the tensor perturbation in the strong-gravity limit.
While the scalar perturbation is affected mainly through the background scalar field,
the tensor perturbation exhibits a direct gravity effect as usual.
In this sense, the resulting form will not be much different from that in the weak-gravity limit
investigated in Ref.~\cite{Cho:2014ija},
except for the values of scales.

We introduce the tensor perturbation fields $\gamma_{ij}$ and $h_{ij}$ as
\begin{align}
ds_q^2 &= -X^2 d\eta^2 +Y^2\left(\delta_{ij}+\gamma_{ij}\right)dx^i dx^j
= Y^2 \left[ -d\tau^2 +\left(\delta_{ij}+\gamma_{ij}\right)dx^i dx^j \right],\label{eqmunu} \\
ds_g^2 &= a^2 \left[ -d\eta^2 +\left(\delta_{ij} +h_{ij}\right) dx^i dx^j \right],\label{gmunu}
\end{align}
where $\tau$ is the conformal time for the auxiliary metric.\footnote{Compared with the case
of the scalar perturbation,
$\tau$ is the same at the near-MPS stage.
At the attractor stage, $\tau$ is different by a factor only in the sub-leading order.
In the leading order, we have $d\tau \approx d\eta$ at the attractor stage
for both cases because $z\approx 1$.}
We impose the transverse and traceless conditions
on both $h_{ij}$ and $\gamma_{ij}$, i.e.,
${\partial}_{i}h^{ij}={\partial}_{i}\gamma^{ij}=0$ and $h=\gamma=0$.
From Eq.~\eqref{eom1}, one then gets $\gamma_{ij} = h_{ij}$ \cite{EscamillaRivera:2012vz},
and also
\be\label{Y}
X=(1+\kappa\rho_0)^{-1/4} (1-\kappa p_0)^{3/4}a,
\qquad
Y=(1+\kappa\rho_0)^{1/4} (1-\kappa p_0)^{1/4}a.
\ee
The Fourier mode is defined by
\be
h_{ij}(\eta,\vec{x}) =\sum_{\sigma = +,-}
\int \frac{d^3k}{(2\pi)^{3/2}} \;
h_{\sigma} (\eta,\vec{k}) \;
\epsilon^{\sigma}_{ij}(\vec{k}) \;
e^{i\vec k\cdot\vec x},
\ee
where $\epsilon^{\sigma}_{ij}$ represents the polarization tensor.
Then, from Eq.~\eqref{eom2} the perturbation equation is given by
\be\label{mueq}
\ddot\mu_\sigma + \left( k^2 -\frac{\ddot Y}{Y} \right)  \mu_\sigma = 0,
\ee
where we introduced a canonical field defined by $\mu_\sigma \equiv (Y/2)h_\sigma$.

As it was mentioned earlier,
the feature of the tensor perturbation is not altered
at the near-MPS stage.
From Ref.~\cite{Cho:2014ija},
$Y\approx (2/3)^{1/4}a_0^{3/2} \sqrt{\tau}$ and
the solution to Eq.~\eqref{mueq} was obtained as
\be\label{mutau}
\mu_{\rm MPS}(\tau)  \approx \sqrt{\tau}\left[c J_0(k\tau)
+ \left( R-i\frac{\pi}{4c}\right)Y_0(k\tau) \right],
\ee
where $c$ and $R$ are the same as in Eq.~\eqref{cR}
with the initial condition imposed in the same way.

At the attractor stage in the strong-gravity limit,
$Y$ in Eq.~\eqref{Y} is approximated as
\be\label{Yatt}
Y \approx (1+\kappa\rho_0)^{1/2}a
\approx \left(1+\frac{1}{2}\kappa m^2\varphi_0^2\right)^{1/2}a
\equiv Y_0a,
\ee
where we used the first slow-roll condition, $\hat{\varphi}_0^2 \ll V$.
In Eq.~\eqref{mueq}, since $d\tau \approx d\eta$,
we observe that the role of the scale factor $a$
for the tensor perturbation in GR is replaced by $Y$.
If $\ddot{Y}/Y \approx \ddot{a}/a$, the mode function will be approximately the same as that in GR.
From Eq.~\eqref{Y}, we get
\begin{align}
\frac{\ddot{Y}}{Y} &= \frac{\ddot{Y_0}}{Y_0} +2\frac{\dot{Y_0}}{Y_0}\frac{\dot{a}}{a}
+\frac{\ddot{a}}{a} \\
&\approx \left( 1-\frac{\kappa m^2}{1+\kappa m^2\varphi_0^2/2} \right)\frac{\ddot{a}}{a}
-\left( \frac{\kappa m^2}{1+\kappa m^2\varphi_0^2/2} \right)
\left( 1+ \frac{\kappa m^2}{1+\kappa m^2\varphi_0^2/2} \right)
\left( \frac{\dot{a}}{a} \right)^2 .\label{ddotY}
\end{align}
If $\kappa m^2 \ll 1$, we have $\ddot{Y}/Y \approx \ddot{a}/a$.
If $\kappa m^2 \gg 1$, Eq.~\eqref{ddotY} is approximated as
\be
\frac{\ddot{Y}}{Y} \approx \left( 1-\frac{2}{\varphi_0^2} \right) \frac{\ddot{a}}{a}
- \frac{2}{\varphi_0^2}\left( 1+ \frac{2}{\varphi_0^2} \right) \left( \frac{\dot{a}}{a} \right)^2.
\ee
Except near the end of the attractor stage,
$\varphi_0$ is quite larger than unity,
and one can still use the approximation $\ddot{Y}/Y \approx \ddot{a}/a$
in the strong-gravity limit.
Then Eq.~\eqref{mueq} becomes
\be\label{mueqATT}
\ddot\mu_\sigma + \left[ k^2 -\frac{2}{(\tau-\tau_0)^2} \right] \mu_\sigma  \approx 0,
\ee
and the mode solution is given by
\be\label{muATT}
\mu_{\rm ATT}(\tau)  \approx A_1 \left[ \cos k(\tau-\tau_0) - \frac{\sin k(\tau-\tau_0)}{k(\tau-\tau_0)} \right]
+A_2 \left[ \sin k(\tau-\tau_0) + \frac{\cos k(\tau-\tau_0)}{k(\tau-\tau_0)} \right].
\ee

Now let us evaluate the tensor power spectrum.
Comparing the solutions for the tensor and the scalar perturbations,
the near-MPS solution $\mu_{\rm MPS}$ in Eq.~\eqref{mutau}
is exactly the same form with $Q_{\rm MPS}$ in Eq.~\eqref{Qsol3},
and the attractor solution $\mu_{\rm ATT}$ in Eq.~\eqref{muATT}
with $Q_{\rm ATT}$ in Eq.~\eqref{muATTmt1}.
If we assume that the tensor perturbation evolves adiabatically,
we can use the same WKB solution also for the tensor perturbation
at the intermediate stage.
Then the solution matching is exactly the same,
and the power spectrum is obtained exactly in the same manner.
The power spectrum at the horizon-crossing is
\begin{align}
P_{\rm T}
&= \frac{k^3}{2\pi^2} h_\sigma^2
= \frac{2k^3}{ \pi^2}\left| \frac{\mu_{\rm ATT}}{Y} \right|^2 \\
&\approx \frac{2}{\pi}\left( c^2+R^2 + \frac{\pi^2}{16c^2} \right)
\times \frac{1}{1+\kappa m^2\varphi_i^2/2}
\times \frac{m^2\varphi_i^2}{6\pi^2} \\
&\equiv D_k \times E_\kappa^{\rm T} \times P_{\rm T}^{\rm GR}.\label{PT}
\end{align}
Here, $P_{\rm T}^{\rm GR} \equiv m^2\varphi_i^2/6\pi^2$ is the spectrum in GR,
$E_\kappa^{\rm T} \equiv 1/(1+\kappa m^2\varphi_i^2/2)$ is the EiBI correction,
and $D_k$ is the same with that in the scalar spectrum.
The result is the same with that in the weak-gravity limit in Ref.~\cite{Cho:2014xaa}.
The difference is that the value of the spectrum can be suppressed almost down to zero
because $\kappa m^2\gg1$  in the strong-gravity limit.

From Eqs.~\eqref{EkS} and \eqref{PT}, we can evaluate the tensor-to-scalar ratio,
\be
r= \frac{P_{\rm T}}{P_{\cal R}}
\approx \frac{E_\kappa^{\rm T} \times P_{\rm T}^{\rm GR}}{E_{\varphi_i}^{\rm S} \times P_{\cal R}^{\rm GR}}
\approx \frac{(1-2/3\varphi_i^2)^{2}}{(1-2/\varphi_i^2)^2 (1+\kappa m^2\varphi_i^2/2)}  \; r^{\rm GR}
\approx \frac{1+8/3\varphi_i^2}{1+\kappa m^2\varphi_i^2/2} \; r^{\rm GR}.
\ee
The EiBI correction for the scalar perturbation reduces the value of $P_{\cal R}$,
but is very tiny.
The EiBI correction for the tensor perturbation reduces the value of $P_{\rm T}$
almost down to zero.
Therefore, the resulting value of the tensor-to-scalar ratio can be suppressed
almost down to zero in the strong-gravity limit.
This is a distinct difference from the chaotic inflation model in GR
which predicts a large value, $r^{\rm GR} \sim 0.131$.
Our result is consistent with the recent analyses by PLANCK collaboration
which predict $r_{0.05} < 0.12$ \cite{Ade:2015tva}
and $r_{0.002} < 0.09$ \cite{Adam:2015rua}.

\begin{figure}[btph]
\begin{center}
\includegraphics[width=.5\linewidth,origin=tl]{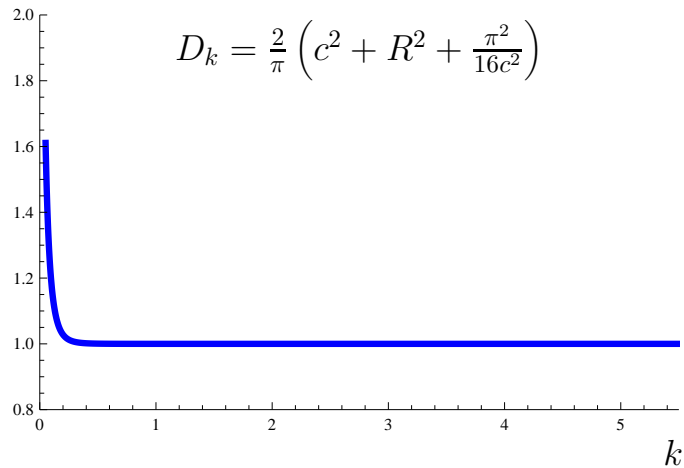}
\end{center}
\caption{Numerical plot of $D_k$ for
$a_0=1$, $\psi_0=-10^{-12}$, $m=10^{-5}$, and $\kappa=10^{13}$.
The value of $\tau_*$ is taken as the lower bound from Eq.~\eqref{lplanck},
and is $k$-dependent.
(In Fig.~1 in Ref.~\cite{Cho:2014xaa}, it was fixed as $\tau_*=1$ for all $k$ modes.)
For the parameters used in this plot, $\tau_*\sim {\cal O}(1)$.
Even with the $k$-dependence in $\tau_*$, the shape of $D_k$ is not changed much,
and still exhibits a peculiar rise for $k\tau_* <1$.
It approaches unity for $k\tau_* >1$.}
\label{FIG1}
\end{figure}

\section{Conclusions}
In this paper, we investigated the primordial density perturbation
in the strong EiBI-gravity limit ($\kappa \gg m^{-2}$).
This is the completion of the series-work on the primordial density perturbation
in EiBI inflation in Refs.~\cite{Cho:2014ija,Cho:2014jta,Cho:2014xaa}.
The scalar and the tensor perturbations are produced at the near-MPS stage.
Imposing the minimum-energy condition for perturbations
at the production moment $\tau_*$
for the wavelength scale $\lambda_{\rm phys} \gtrsim l_p$,
we evaluated the power spectra when the perturbations exit the horizon
near the beginning of the inflationary attractor stage.

In the strong-gravity limit,
the first-order EiBI correction for the scalar perturbation
comes from the background scalar-field configuration.
The correction to the power spectrum lowers its value,
but it is very tiny.

The EiBI correction for the tensor perturbation comes directly from
the gravity effect through the EiBI theory parameter $\kappa$.
The formula for the power spectrum is the same regardless of the strength of gravity.
However, the value of the spectrum can be suppressed significantly
almost down to zero.

The resulting tensor-to-scalar ratio is suppressed almost down to zero.
This supports the recent data analyses of the PLANCK observational data.
In this respect, while the $\varphi^2$ inflation model in GR is ruled out
by the observational result,
it is very viable in EiBI gravity.
When we remind that the contribution to the total power spectrum
mainly comes from the scalar perturbation,
the change in the value of the total power spectrum
is within the observation bound.
In order to check the viability of the inflation model,
another key point is the spectral index.
The value of the spectral index that the chaotic inflation model in GR predicts,
fits the observational data very nicely.
The recent investigation shows that the EiBI correction to the spectral index
is the second order of slow-roll parameters for the scalar perturbation,
and the first order of the tensor perturbation \cite{ChoGong}.
Therefore, it is well within the observational fit.

In this work, we considered the strong-gravity limit, $\kappa \gg m^{-2} \sim 10^{10}$.
From the star formation study in Refs. \cite{Avelino:2012ge,Pani:2011mg,Pani:2012qb},
the value of the parameter $\kappa$ is very mildly constrained as
$\kappa < 10^{-2} m^5 kg^{-1}s^{-2} \sim 10^{77}$ in Planck unit.
Therefore, our strong limit resides well within the constraint.

After the perturbations cross in the horizon during the radiation/matter-dominated epoch,
they evolve in a very similar way as in GR.
From the study of perfect fluid in EiBI cosmology \cite{Cho:2012vg},
the background Universe settles down to the standard evolution in GR
when $\kappa\rho \ll 1$.
Considering that the energy scale at the beginning of the radiation-dominated epoch
is $\rho \gtrsim 1 {\rm MeV} \sim 10^{-22}$,
a wide range of the value of $\kappa$ in the strong limit
satisfies the condition $\kappa\rho \ll 1$.

\section*{Acknowledgement}
This work was supported in part by the grant from the National Research Foundation
funded by the Korean government, No. NRF-2012R1A1A2006136.


\begin{thebibliography}{99}

\bibitem{Ade:2014xna}
  P.~A.~R.~Ade {\it et al.}  [BICEP2 Collaboration],
  Phys.\ Rev.\ Lett.\  {\bf 112}, 241101 (2014)  [arXiv:1403.3985 [astro-ph.CO]].

\bibitem{Ade:2015tva}
  P.~A.~R.~Ade {\it et al.}  [BICEP2 and Planck Collaborations],
  [arXiv:1502.00612 [astro-ph.CO]].

\bibitem{Adam:2015rua}
  R.~Adam {\it et al.}  [Planck Collaboration],
  arXiv:1502.01582 [astro-ph.CO].

\bibitem{Linde:1983gd}
  A.~D.~Linde,
  Phys.\ Lett.\ B {\bf 129} (1983) 177.

\bibitem{Cho:2013pea}
  I.~Cho, H.~-C.~Kim and T.~Moon,
  Phys.\  Rev.\  Lett 111, {\bf 071301} (2013)
  [arXiv:1305.2020 [gr-qc]].

\bibitem{Banados:2010ix}
  M.~Banados and P.~G.~Ferreira,
  Phys.\ Rev.\ Lett.\  {\bf 105}, 011101 (2010)
  [arXiv:1006.1769 [astro-ph.CO]].

\bibitem{Cho:2014jta}
  I.~Cho and N.~K.~Singh,
  Eur.\ Phys.\ J.\ C {\bf 74}, no. 11, 3155 (2014)  [arXiv:1408.2652 [gr-qc]].

\bibitem{Cho:2014xaa}
  I.~Cho and N.~K.~Singh,
  Eur.\ Phys.\ J.\ C {\bf 75}, no. 6, 240 (2015)  [arXiv:1412.6344 [gr-qc]].  

\bibitem{Cho:2014ija}
  I.~Cho and H.~-C.~Kim,
  Phys.\ Rev.\ D {\bf 90}, 024063 (2014)  [arXiv:1404.6081 [gr-qc]].

\bibitem{Lagos:2013aua}
  M.~Lagos, M.~Banados, P.~G.~Ferreira and S.~Garcia-Saenz,
  Phys.\ Rev.\ D {\bf 89}, 024034 (2014)  [arXiv:1311.3828 [gr-qc]].

\bibitem{Scargill:2012kg}
  J.~H.~C.~Scargill, M.~Banados and P.~G.~Ferreira,
  Phys.\ Rev.\ D {\bf 86}, 103533 (2012)  [arXiv:1210.1521 [astro-ph.CO]].

\bibitem{EscamillaRivera:2012vz}
  C.~Escamilla-Rivera, M.~Banados and P.~G.~Ferreira,
  Phys.\ Rev.\ D {\bf 85}, 087302 (2012)
  [arXiv:1204.1691 [gr-qc]].

\bibitem{ChoGong}
  I.~Cho and J.~O.~Gong,
  ``Spectral index in EiBI inflation,''
  {\it in preparation}.

\bibitem{Pani:2011mg}
  P.~Pani, V.~Cardoso and T.~Delsate,
  Phys.\ Rev.\ Lett.\  {\bf 107}, 031101 (2011)
  [arXiv:1106.3569 [gr-qc]].

\bibitem{Pani:2012qb}
  P.~Pani, T.~Delsate and V.~Cardoso,
  Phys.\ Rev.\ D {\bf 85}, 084020 (2012)
  [arXiv:1201.2814 [gr-qc]].

\bibitem{Avelino:2012ge}
  P.~P.~Avelino,
  Phys.\ Rev.\ D {\bf 85}, 104053 (2012)  [arXiv:1201.2544 [astro-ph.CO]].

\bibitem{Cho:2012vg}
  I.~Cho, H.~-C.~Kim and T.~Moon,
  Phys.\ Rev.\ D {\bf 86}, 084018 (2012)  [arXiv:1208.2146 [gr-qc]].  




\end{thebibliography}
\end{document}